\documentclass[twocolumn,twocolappendix]{aastex631}

\usepackage{multirow}
\usepackage{xspace}
\usepackage{enumitem}
\usepackage[T1]{fontenc}
\usepackage{soul}

\newcommand{\gm}{\ensuremath{\gamma}}
\newcommand{\fermi}{\textit{Fermi}\xspace}
\newcommand{\swift}{\textit{Swift}\xspace}

\newcommand{\xmm}{\textit{XMM-Newton}\xspace}
\newcommand{\nustar}{\textit{NuSTAR}\xspace}
\newcommand{\txs}{TXS\,1508+572\xspace}

\begin{document}

\title{A gamma-ray flare from \txs: characterizing the jet of a $z=4.31$ blazar in the early Universe}

\email{gokus@wustl.edu}

\author[0000-0002-5726-5216]{Andrea Gokus}
\affiliation{Department of Physics \& McDonnell Center for the Space Sciences, Washington University in St. Louis, One Brookings Drive, St. Louis, MO 63130, USA}

\author[0000-0002-8434-5692]{Markus B\"ottcher}
\affiliation{Centre for Space Research, North-West University, Potchefstroom 2520, South Africa}

\author[0000-0002-1853-863X]{Manel Errando}
\affiliation{Department of Physics \& McDonnell Center for the Space Sciences, Washington University in St. Louis, One Brookings Drive, St. Louis, MO 63130, USA}

\author[0000-0002-3092-3506]{Michael Kreter}
\affiliation{Centre for Space Research, North-West University, Potchefstroom 2520, South Africa}

\author[0009-0009-7841-1065]{Jonas He{\ss}d\"orfer}
\affiliation{Julius-Maximilians-Universität Würzburg, Fakultät für Physik und Astronomie, Institut für Theoretische Physik und Astrophysik, Lehrstuhl für Astronomie, Emil-Fischer-Str. 31, D-97074 Würzburg, Germany}

\author[0000-0001-7112-9942]{Florian Eppel}
\affiliation{Julius-Maximilians-Universität Würzburg, Fakultät für Physik und Astronomie, Institut für Theoretische Physik und Astrophysik, Lehrstuhl für Astronomie, Emil-Fischer-Str. 31, D-97074 Würzburg, Germany}

\author[0000-0001-5606-6154]{Matthias Kadler}
\affiliation{Julius-Maximilians-Universität Würzburg, Fakultät für Physik und Astronomie, Institut für Theoretische Physik und Astrophysik, Lehrstuhl für Astronomie, Emil-Fischer-Str. 31, D-97074 Würzburg, Germany}

\author[0000-0002-5083-3663]{Paul~S.~Smith}
\affiliation{Steward Observatory, The University of Arizona, Tucson, AZ 85721, USA}

\author[0009-0006-4186-9978]{Petra Benke}
\affiliation{Max-Planck-Institut für Radioastronomie, Auf dem Hügel 69, D-53121 Bonn, Germany}
\affiliation{Julius-Maximilians-Universität Würzburg, Fakultät für Physik und Astronomie, Institut für Theoretische Physik und Astrophysik, Lehrstuhl für Astronomie, Emil-Fischer-Str. 31, D-97074 Würzburg, Germany}

\author[0000-0002-0694-2459]{Leonid I. Gurvits}
\affiliation{Joint Institute for VLBI ERIC (JIVE), Oude Hoogeveensedijk 4, 7991 PD, Dwingeloo, The Netherlands}
\affiliation{Faculty of Aerospace Engineering, Delft University of Technology, Kluyverweg 1, 2629 HS, Delft, The Netherlands}

\author[0000-0002-4184-9372]{Alex Kraus}
\affiliation{Max-Planck-Institut für Radioastronomie, Auf dem Hügel 69, D-53121 Bonn, Germany}

\author[0000-0001-6088-3819]{Mikhail Lisakov}
\affiliation{Instituto de F\'{i}sica, Pontificia Universidad Cat\'{o}lica de Valpara\'{i}so, Casilla 4059, Valpara\'{i}so, Chile}
\affiliation{Max-Planck-Institut für Radioastronomie, Auf dem Hügel 69, D-53121 Bonn, Germany}

\author[0000-0001-6191-1244]{Felicia McBride}
\affiliation{Department of Physics and Astronomy, Bowdoin College, Brunswick, ME 04011, USA}

\author[0000-0001-9503-4892]{Eduardo Ros}
\affiliation{Max-Planck-Institut für Radioastronomie, Auf dem Hügel 69, D-53121 Bonn, Germany}

\author[0009-0000-4620-2458]{Florian R\"osch}
\affiliation{Julius-Maximilians-Universität Würzburg, Fakultät für Physik und Astronomie, Institut für Theoretische Physik und Astrophysik, Lehrstuhl für Astronomie, Emil-Fischer-Str. 31, D-97074 Würzburg, Germany}
\affiliation{Max-Planck-Institut für Radioastronomie, Auf dem Hügel 69, D-53121 Bonn, Germany}

\author[0000-0003-2065-5410]{J\"orn Wilms}
\affiliation{Remeis Observatory \& Erlangen Centre for Astroparticle Physics, Universit\"at Erlangen-N\"urnberg, Sternwartstr.~7, 96049 Bamberg, Germany}

\begin{abstract}
Blazars can be detected from very large distances due to their high luminosity. However, the detection of \gm-ray emission of blazars beyond $z=3$ has only been confirmed for a small number of sources. Such observations probe the growth of supermassive black holes close to the peak of star formation in the history of galaxy evolution. 
As a result from a continuous monitoring of a sample of 80 $z>3$ blazars with \fermi-LAT, we present the first detection of a \gm-ray flare from the $z=4.31$ blazar \txs.
This source showed high \gm-ray activity from February to August 2022, reaching a peak luminosity comparable to the most luminous flares ever detected with \fermi-LAT. We conducted a multiwavelength observing campaign involving \xmm, \swift, the Effelsberg 100-m radio telescope and the Very Long Baseline Array. In addition, we make use of the monitoring programs by the Zwicky Transient Facility and {\it NEOWISE} at optical and infrared wavelengths, respectively.
We find that the source is particularly variable in the infrared band on daily time scales. The spectral energy distribution collected during our campaign is well described by a one-zone leptonic model, with the \gm-ray flare originating from an increase of external Compton emission as a result of a fresh injection of accelerated electrons.
\end{abstract}


\section{Introduction} \label{sec:intro}

Among jetted active galactic nuclei (AGN), those with their jet pointing towards the Earth are called blazars and appear particularly variable and luminous due to relativistic beaming \citep{urry1995}. Blazars can be broadly classified as either BL Lacertae (BL Lac) objects or flat-spectrum radio quasars (FSRQs), for which the distinction is based on the existence of optical emission lines with an equivalent width of at least 5\AA (FSRQ) or the lack thereof (BL Lacs). 

Due to their extreme luminosities, we are able to detect quasars and blazars at large distances, currently up to $z=7.6$ \citep{wang2021}, which enables glimpses into the early Universe. Among those found at high redshift, some seem to be powered by the heaviest specimens of black holes, exceeding a billion solar masses \citep[e.g.,][]{lobanov2001,ghisellini2010, ackermann2017, belladitta2022,burke2024}.
Given that these extremely massive black holes appear to exist only about one billion years after the Big Bang, the circumstances under which black holes can grow so fast are not yet understood \citep[see, e.g.,][for a recent review]{inayoshi2020}.
While our observations are biased towards finding the most luminous and therefore extreme sources, current estimations for their number densities present a hurdle for the application of our existing models that describe black hole formation and growth \citep[e.g.,][]{johnson2016}.
Some of the suggestions for black hole formation for heavy seeds are remnants of the supernovae of massive first-generation stars \citep{madau2001}, or matter collapsing into SMBHs right away \citep[e.g.,][]{begelman2006}. However, even with black hole seeds of $\sim$\,100~M$_{\odot}$ the time scales needed for SMBHs to grow to $>10^9$\,M$_{\odot}$ are too long even when assuming accretion at the Eddington limit throughout the entire time. 
Work by \citet{volonteri2011} and \citet{ghisellini2013} investigate the role of AGN feedback in relation to black hole growth and propose a scenario in which powerful jets can prompt a higher accretion rate. \citet{alexander2014} have suggested that the first black holes might have been able to grow fast through supra-exponential accretion.
A recent study by \citet{lai2024} argues that SMBHs at $z\sim5$ are only growing slowly, thereby requiring either initial seed masses $>10^8$~M$_{\odot}$ or extremely rapid growth at $z>5$.
Simulations of rapidly spinning black holes accreting above the Eddington limit have revealed that their growth is accompanied by the formation of powerful jets \citep{mckinney2014, sadowski2014}.
Hence, high-redshift ($z>3$) blazars are ideal targets in order to learn more about the growth of SMBHs, while also taking into account that their powerful large-scale jets influence their host galaxies and galaxy clusters, which affects galaxy evolution. 

In order to properly assess the physical properties of these blazars, it is essential to obtain a multiwavelength data set for these objects, in particular, information about the high-energy emission that is needed to measure the full power of the jet.
While at X-ray energies we have been able to probe distances up to $z>6$ 
\citep[e.g.,][]{sbarrato2015, belladitta2020, medvedev2020, sbarrato2022, migliori2023}, we have not been able to do so at \gm-ray energies, even though the Universe is transparent up to 10\,GeV \citep{dominguez2024}. However, due to both the large distance and the shift of the emitted \gm-ray emission to lower energies where the sensitivity of \fermi-LAT decreases, most of the blazars appear to be too faint to be detected.
Studies on high-z blazars are currently limited to a small sample of objects with $z\geq3$. The current version of the AGN catalog based on data from the \fermi Large Area Telescope (LAT) lists 11 \citep{4lac,4lac_dr3}, and other studies have identified three more high-$z$ blazars \citep{liao2018,kreter2020}.

A blazar's broadband spectral energy distribution (SED) consists of two 
broad, non-thermal components, where the low-energy component can be generally attributed to leptonic synchrotron emission, while the high-energy component can be explained by leptonic Inverse Compton processes \citep[e.g.,][]{maraschi1992, dermer1993, sikora1994, blandford1995, bloom1996, boettcher1997, blazejowski2000}, but also additional hadronic interactions such as photon-pion interactions and proton synchrotron emission \citep[e.g.,][]{mannheim1992,mannheim1993, aharonian2000, muecke2001, aharonian2002, muecke2003,Boettcher13}. 
In addition, some blazars feature a thermal component from their accretion disks, which are particularly present in high-z sources for which the emission from the disk is redshifted to 
optical and even infrared wavelengths.
As the entire emission from high-$z$ blazars becomes redshifted, the peak of the high-energy component drifts to MeV energies, which originally coined the term `MeV blazar' \citep{bloemen1995}. In addition, \citet{sikora2002} have also presented an underlying physical distinction from GeV blazars that relates to electron cooling through Comptonization of near-IR emission from the dusty torus in the case of MeV blazars.
As a result, the \gm-ray spectra of these objects are steep \citep{ackermann2017}.

In Section~\ref{sec:campaign}, we outline our strategy of detecting \gm-ray flares from high-$z$ blazars and introduce \txs.
Our data reduction is described in Section~\ref{sec:mwldata}. We compute the \gm-ray luminosity displayed by \txs during its peak in Section~\ref{sec:max_gammaflux} and assess the changes in the \gm-ray, X-ray, and radio spectra in Section~\ref{sec:results_spectra}. 
In addition, our results of the multiwavelength data analysis include a variability study of \txs during the flaring state as well as some long-term variability in Section~\ref{sec:results_variability} and a modeling of the broadband SED in Section~\ref{sec:results_sedmodel}.
Finally, we discuss all our findings and conclude in Section~\ref{sec:discuss_conclude}.
Throughout this paper, we assume a flat cosmology following \citet{planck_collab2016}, which reported H$_0$=67.8 km/s, $\Omega_{\lambda}=0.692$, and $\Omega_M=0.308$. With those parameters, \txs has a luminosity distance of $\sim 40$\,Gpc.

\section{Monitoring campaign}\label{sec:campaign}
While several studies have been collecting multiwavelength data for MeV blazars and have modeled their SEDs using non- or semi-contemporaneous data \citep[e.g.,][]{paliya2016, ackermann2017, marcotulli2020}, some have reported on the analysis of flares from high-$z$ blazars using quasi-simultaneous data sets \citep[albeit some sources with $2<z<3$;][]{orienti2014, paliya2019, liao2019}.
In order to be able to catch a high-$z$ blazar during a \gm-ray flare, we use \fermi-LAT's all-sky monitoring capabilities and set up a \gm-ray monitoring program for 80 sources for which the fifth version of the ROMA-BZCAT \citep[5BZCAT][]{bzcat_1stedition, bzcat_5thedition} lists $z>3$.
Our method relies on a significant detection (i.e., TS\,$\geq25$, where the test statistic $\mathrm{TS}=2\Delta\mathrm{log}(\mathcal{L})$ compares the likelihood with and without a source; see Sect.~\ref{sec:data_gamma} for details) for 30-day average time bins, for which we perform a daily check. Using 30-day average time bins has been proven successful in the search for high-$z$ blazars using archival data \citep{kreter2020}.

In order to obtain a simultaneous multiwavelength data set upon flare detection, we set up follow-up observations using multiple facilities (100-m Effelsberg radio telescope, \xmm, \swift) and also make use of exisiting all-sky survey facilities such as ZTF and {\it NEOWISE}.

In this work, we report on the first \gm-ray flare detected for a $z>4$ blazar, \txs, which occured in February 2022 and marked the beginning of a high-activity phase lasting roughly 6 months.

\subsection{\txs}
At a redshift of $z=4.31$ \citep[][but note that the first report of its redshift at $z=4.30$ was done by \citet{hook1995}]{schneider2007}, the blazar \txs is the third most distant \gm-ray source detected with \fermi-LAT data as reported by \citet{liao2018} and \citet{kreter2020} \citep[but note that in the most recent release of the \fermi-LAT AGN Catalog (4LAC-DR3) by][it is still listed as the most distant \gm-ray emitter]{4lac_dr3}. As a \gm-ray emitter, the source is known as 4FGL\,J1510.1+5702, but historically has been referred to as GB\,1508+5714.
The source was first studied at X-ray energies with {\it Einstein} data \citep{mathur1995}, and subsequently studied by {\it ASCA} \citep{moran1997}. 
Using {\it Chandra}'s high-resolution ACIS detector, \citet{siemiginowska2003} and \citet{yuan2003} independently reported on the identification of an X-ray jet seen as extended emission from the source.
The first very long baseline interferometry (VLBI) high-resolution image, taken with the European VLBI Network (EVN) at 5\,GHz with the angular resolution of $\sim 5$\, mas in 1995, did not reveal any extended radio emission \citep{frey1997}. 
However, Very Large Array (VLA) snapshot observations taken at 1.4\,GHz were able to detect a low-brightness jet extending to $\sim 2.5$\,arcsec toward south-west \citep{cheung2004}.
\citet{osullivan2011} reported an optically thin jet component at 5 GHz and 8.4 GHz roughly 2 mas to the south of the core based on global VLBI polarimetry observations.
A recent observation of the source with LOFAR revealed emission from the counter-jet seen at 144\,MHz with sub-arcsecond spatial resolution \citep{kappes2022}. 
On milli-arcsecond scales, \citet{titov2023} monitored \txs over four years at 2.3 and 8.4\,GHz in order to study the apparent absolute astrometric proper motion and detected a jet proper motion of $0.117\pm 0.078$~mas/yr.

In 2017, the object was confirmed as a \gm-ray emitter \citep{ackermann2017}, and variability on monthly time scales has been detected \citep{li2018}. A period of enhanced brightness in both the \gm-ray and the optical band has been reported by \citet{liao2020}.

On 4 February 2022, we detected a \gm-ray flare of \txs through our \fermi-LAT monitoring pipeline and found that the source had significantly brightened compared to the flux reported in the \fermi-LAT Fourth Source Catalog \citep[4FGL;][]{4fgl,gokus2022}. We coordinated a multiwavelength campaign to obtain simultaneous data during the flaring state, and in addition launched a VLBI monitoring campaign using the Very Long Baseline Array (VLBA) and the 100-m Effelsberg radio telescope, whose results we report in \citet{benke2024}.

\section{Multiwavelength data}\label{sec:mwldata}
\begin{figure*}
    \centering
    \includegraphics[width=.99\textwidth]{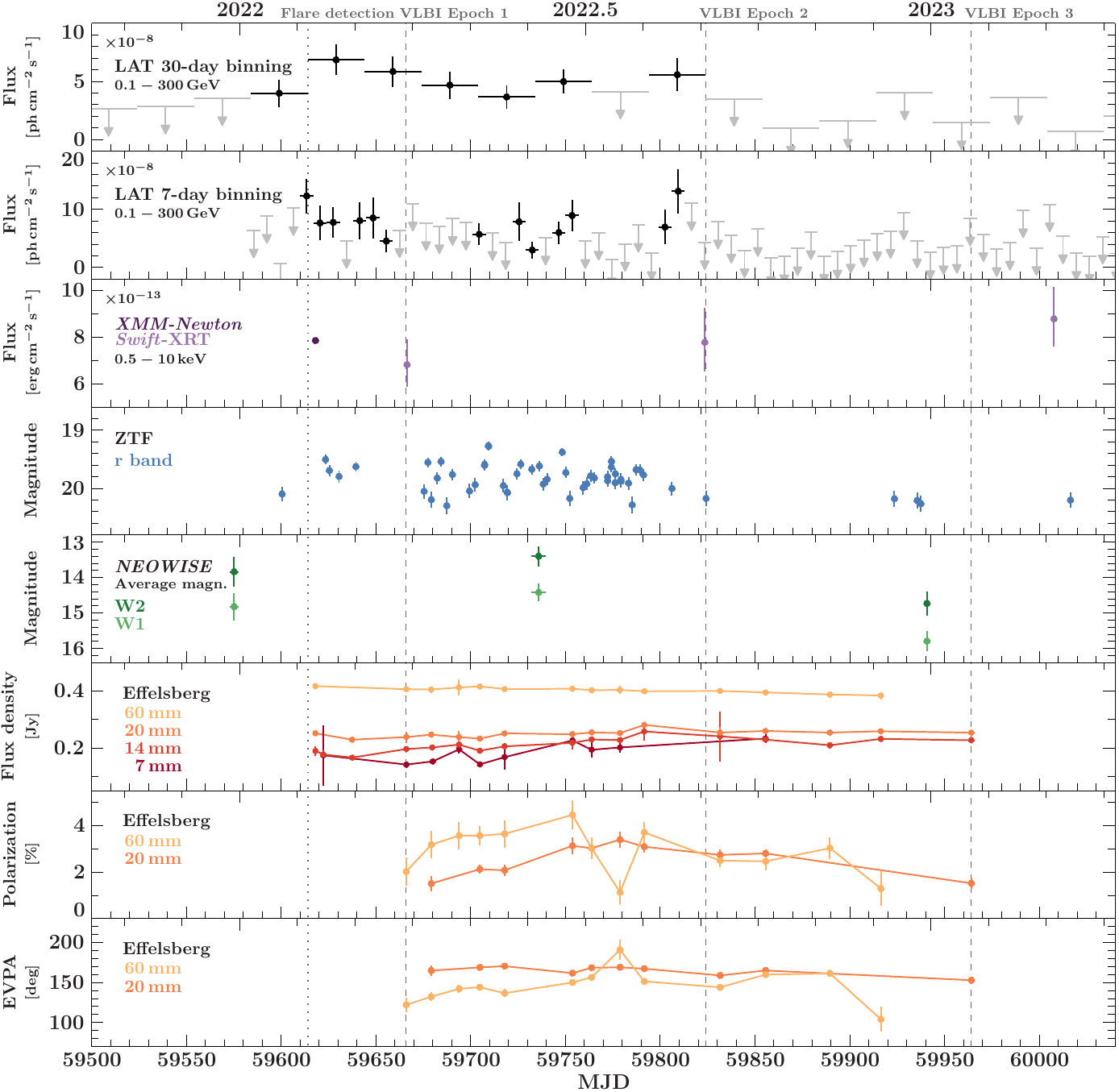}
    \caption{Multiwavelength light curve of \txs covering a time span from the end of 2021 to the beginning of 2023. The panels from top to bottom show the fluxes, magnitudes, or flux densities in different, decreasing energy ranges, starting with \gm-rays at the top. The two bottom panels illustrate changes in the single-dish radio polarization and EVPA, respectively, measured at 60\,mm and 20\,mm. A dotted line marks the detection of the flaring state of \txs, while the three dashed lines mark the VLBI observations presented in \citet{benke2024}.}
    \label{fig:MWL_LC}
\end{figure*}

In this section, we report on the broadband data extraction and analysis, and show the fluxes, magnitudes, and flux densities that have been compiled in light curves for different energy bands in Fig.~\ref{fig:MWL_LC}.

\subsection{\fermi-LAT}\label{sec:data_gamma}
The Large Area Telescope (LAT) onboard the \fermi satellite \citep{fermi_instr} has monitored the entire sky since August 2008, and has detected 3814 AGN in 12 years \citep{4lac,4lac_dr3}, with the large majority being blazars. 
We extract the LAT data using ScienceTools Version 1.2.23 and fermipy 0.20.0 \citep{fermipy}, following the standard data reduction process.
Events with an energy between 100\,MeV and 300\,GeV are selected within a region of interest (ROI) of 15 degrees around the \gm-ray source coordinates given in the \fermi-LAT Fourth Source Catalog Paper \citep[4FGL;][]{4fgl}, and filtered by selecting SOURCE class events, and those fulfilling \texttt{DATA\_QUAL$>$0} and \texttt{LAT\_CONFIG==1}. In addition, we use only events that enter the instrument with a maximum zenith angle of $90^\circ$ to avoid contamination by \gm-rays from Earth-limb effects.
To model the background, we use \texttt{gll\_iem\_v07} as the Galactic diffuse model, and \texttt{iso\_P8R3\_SOURCE\_V2\_v1} to account for the isotropic diffusion emission. Furthermore, we use the post-launch instrument response function \texttt{P8R3\_SOURCE\_V2}.
In addition to the background, we include all known \gm-ray sources listed in the 4FGL that lie within $20^{\circ}$ of TXS\,1508+572, and use the spectral type listed in the 4FGL as model for each source. In the case of our target source, this model is represented by a power law.
Optimization of our model parameters is done via a maximum likelihood analysis, for which the significance of the \gm-ray emission for each source is determined via the test statistic $\mathrm{TS}=2\Delta\mathrm{log}(\mathcal{L})$. The likelihood function $\mathcal{L}$ describes the difference between a model with and without a source at the given coordinates \citep{mattox}.
After an initial fit of all components within the ROI, we remove point sources that are found with TS $<4$ or TS = NaN as their background contribution is minimal to non-existent. By checking the resulting TS maps, we ensure that no excess emission is left over after the fitting procedure.

The \gm-ray spectra of \txs are created for the quiescent and flaring states, which cover a time range of MJD 55197 to 59215 (2010-01-01 to 2021-01-01), and MJD 59610 to 59624 (2022-01-31 to 2022-02-14), respectively.
In the spectral fits, we keep spectral and normalization parameters free for our target source and all sources within $3^{\circ}$ of it, which are initially five sources in the first fit that were all kept for the quiescent state, but removed for the fit of the flaring state. 
For sources within $5^{\circ}$, or with $\mathrm{TS}>500$, we leave the normalization free to vary. Galactic and isotropic diffuse emissions are also kept free during the modeling.

Moreover, we compute a light curve from MJD 59434 to 60094 (2021--08--08 to 2023--05--30) with 30-day binning, and MJD 59582 to MJD 60051 (2022--01--03 to 2023--04--17) with 7-day binning, which are shown in the top two panels of Fig.~\ref{fig:MWL_LC} for the 30-day and 7-day binning, respectively.
The 30-day binning is chosen such that the first TS$\geq25$ bin coincides with the detection of increased activity with our pipeline, i.e., on 4 February 2022.
We keep the normalization free for all sources within $3^{\circ}$ of TXS\,1508+572, as well as sources with $\mathrm{TS}>500$ over the entire time range. In addition, we keep the spectral index of our target source as a free parameter. All diffuse emission is kept frozen to the best-fit parameters determined in the modeling of the entire time range.
We display the $2\sigma$ upper limit for bins with $\mathrm{TS}<25$ for the 30-day binned light curve, and $\mathrm{TS}<9$ for the 7-day binned light curve, which equates to a detection significance of $\sim5\sigma$ and $\sim3\sigma$, respectively.
The uncertainties of the flux bins are given as $1\sigma$ uncertainties.

\subsection{X-ray data}\label{sec:xraydata}
We model all X-ray data using the Interactive Spectral Interpretation System \citep[ISIS, Version 1.6.2-51; ][]{isis} in order to determine spectral parameters and the source flux, and utilize the C-statistics \citep{cash}. Uncertainties are given at the $1\sigma$ confidence level. The absorption by the interstellar medium is based on the \texttt{vern} cross sections \citep{vern} and the \texttt{wilm} abundances \citep{wilms_tbabs}. 
For each spectrum, we 
use an absorbed, pegged power law (\texttt{tbabs*pegpwrlw}), except for a simultaneous fit of \xmm and \nustar spectra, which we utilize to test for a spectral break.
We freeze the Galactic hydrogen absorption to the value from the HI 4$\pi$ survey \citep[HI4PI; ][]{HI4PI}, which is $N_\mathrm{H,Gal}=1.55\times10^{20}\,\mathrm{cm}^{-2}$.
The results from our X-ray spectral fits are presented in Sect.~\ref{sec:results_spectra}.
The fluxes measured with \xmm and \swift and obtained from the respective best fits are shown in the third panel from the top in Fig.~\ref{fig:MWL_LC}.

\subsubsection{\xmm}\label{sec:xmm_extract}
We obtained a target-of-opportunity (ToO) observation with the \xmm observatory \citep{jansen2001} to follow up the flare as soon as possible after the detection of the \gm-ray flare. \txs was observed on 8 February 2022 (ObsID 0910390101) for 85\,ks in the \textit{Small Window} mode with the European Photon Imaging Camera (EPIC) with the pn \citep{epic-pn} and both MOS detectors \citep{epic-mos}. 
In addition, we extract the data from an archival observation performed on 11 May 2002 (ObsID 0111260201), which was taken in \textit{Full Window} mode with an exposure time of 24\,ks. For the archival data, we were only able to obtain spectra from the EPIC-pn and EPIC-MOS2 detector since the source region coincides with a CCD gap for the EPIC-MOS1 detector.
We extract spectra for both observations using the \xmm Science Analysis Software (SAS, Version 20.0.0) by using the standard methods in order to process the observation data files and produce calibrated event lists and images. 
The circular source regions in each detector are centered on the point source at the source coordinates with a radius of 35 arcsec for the EPIC-pn detector as well as the EPIC-MOS2 detector for ObsID 0111260201, and a radius of 27.5 arcsec for all other EPIC-MOS detectors. The background regions are circles positioned in a source free region at sufficient distance from the target source, and have a radius of 60 arcsec for all detectors except for the EPIC-MOS2 detector in ObsID 0111260201, for which we have chosen 100 arcsec.
We extract single and double events for the data taken with EPIC-pn, and all events for data taken with the EPIC-MOS detectors. In both observations, pile-up is negligible. 
We extract light curves of the observation in three different energy bands using the EPIC-pn detector: for a full energy range from 0.3\,keV to 10\,keV, as well as for a soft band from 0.3\,keV to 2\,keV and for a hard band from 2\,keV to 10\,keV.
We find that at the beginning and end of the observation, the data are impacted by severe background flaring and exclude those time ranges from any data analyses. For the remaining time range, we compute a subtracted light curve by subtracting the background light curve from the light curve obtained for the source region. To correct for detector and scaling effects, we only work with light curves produced with the XMMSAS function \texttt{epiclccorr}. Note that we also exclude all bins with a fractional exposure below 65\%.

\subsubsection{Neil Gehrels \swift Observatory}
We requested several observations with the {\it Neil Gehrels Swift Observatory} \citep{swift} over the course of a year, to gather contemporaneous data to the VLBI monitoring with VLBA and the 100-m Effelsberg radio telescope \citep{benke2024}.
In addition to the observations obtained in relation to the \gm-ray flare, an archival observation of TXS\,1508+572 exists, which was taken simultaneous to a {\it NuSTAR} observation in 2017. We use those data for creating a \gm-ray quiescent-state SED of the source.
The \swift-XRT data are extracted using the standard procedures with the \texttt{xrtpipeline} (Version 0.13.7) with HEASOFT 6.30.
The source spectra are compiled from the source coordinates (R.A. = 227.5114780$^{\circ}$, Decl. = 57.0447472$^{\circ}$) and a radius of 30\,arcsec, while the background is created using an annulus centered on the same coordinates, with an inner radius of 40\,arcsec and an outer radius of 150\,arcsec.

We do not detect TXS\,1508+572 with the optical/UV telescope onboard \swift.

\subsubsection{\nustar}
We utilize an archival \nustar observation taken on 30 April 2017 (ObsID 60201013002), with the exposure time of 73\,ks, to obtain hard X-ray data for the quiescent state SED. We extract the data using the standard methods with NUSTARDAS (Version 2.1.2) and CALBD version 20230124.
Using nupipeline (version 0.4.9) we reduce data from both Focal Plane Modules A and B (FPMA and FPMB), and create calibrated event lists and images. We choose a circular region with a radius of 50 arcsec at R.A. = 227.5131365$^{\circ}$, Decl. = 57.0465406$^{\circ}$ and R.A. = 227.5138372$^{\circ}$, Decl. = 57.0473028$^{\circ}$ for the source region in FPMA and FPMB, respectively. The background regions are chosen to be circular as well with a radius of 160 arcsec, and centered on R.A. = 227.4806152$^{\circ}$, Decl. = 57.1443941$^{\circ}$, and R.A. = 227.4856466$^{\circ}$, Decl. = 57.1484921$^{\circ}$ for FPMA and FPMB, respectively.
After the data extraction, the resulting spectra have an exposure time of $\sim37$\,ks.

\subsection{Optical \& Infrared archival data}
Through Lyman-$\alpha$ ($\lambda$1215.67\,\AA) absorption, the intergalactic medium affects the optical emission from distant sources \citep[e.g.,][]{gunn1965}. With a redshift of $z=4.31$, the optical light of \txs is absorbed at wavelengths starting at $6443$\,\AA, which falls in the middle of the range of red filters.

The optical and infrared data are displayed in two middle panels of Fig.~\ref{fig:MWL_LC}.

\subsubsection{XMM-OM}
The \textit{Optical Monitor} (OM) onboard \xmm observed \txs simultaneous to the X-ray band, and took images in the V, U, W1, W2, and M2 filters with a net exposure of 12\,ks, 12\,ks, 16\,ks, 20\,ks, and 16\,ks, respectively.
We extract the photometric information using \texttt{omichain}, which is part of the \xmm SAS.
The source is only detected in the V, U, and W1 band. However, all bands are affected by Lyman-$\alpha$ absorption. We include these photometric measurements in the broadband SED for display purposes.

\subsubsection{Zwicky Transient Facility}
We obtained optical data from the Zwicky Transient Facility \citep[ZTF;][]{ztf_mainpaper,ztf_dataarchive,zwicky_archive_doi} survey through their public data release (DR18). The data consist of photometric measurements in the \textit{gri} filter system. The data taken in the \textit{g}-band are fully affected by absorption in the Lyman-$\alpha$ forest, and are not taken into account in this work.
The ZTF survey data for \txs shown in this work cover a time range from March 2018 through February 2023, with gaps in between epochs of dense monitoring. 

\subsubsection{NEOWISE}
At near-infrared wavelength, we use public data from the {\it NEOWISE} mission \citep{neowise,neowise_public_archive}, in particular the 2024 data release, which includes data from 25 December 2013 up to 6 June 2023.
Data are available for two wavelength bands, which are 3.4 and 4.6 $\mu\mathrm{m}$, and we have initially selected all data available within 5 arcseconds around the source coordinates from the online database\footnote{\href{https://irsa.ipac.caltech.edu/cgi-bin/Gator/nph-dd}{https://irsa.ipac.caltech.edu/cgi-bin/Gator/nph-dd}}.
For the data selection, we have applied the following criteria (see description in \url{anjum2020}, which follows \citet{rakshit2019}):
\begin{enumerate}[noitemsep]
    \item The fit quality given as the reduced $\chi^2$ (\texttt{w1rchi2} / \texttt{w2rchi2}) is less than 5 in both photometric bands.
    \item The number of components used to perform a profile fit (\texttt{nb}) of the point spread function is $<3$.
    \item The single-exposure images exhibit the best quality (\texttt{qi\_fact}\,$=1$) and are unaffected by known artifacts (\texttt{cc\_flags}\,='0000') and not actively de-blended (\texttt{na}\,$=0$).
\end{enumerate}
Additionally, we only take into account frames with the highest quality (\texttt{qual\_frame}\,$=10$).
In our analysis, we only use measured magnitudes, that is, magnitudes for which an uncertainty is given, but no upper limits.
We note that one W2 band observation taken in December 2019 shows a magnitude of $\sim$\,12 mag for \txs, but for the simultaneous observation in the W1 band, no significant increase is present. Hence, even though data quality is good and an uncertainty is available for that particular observation, we exclude it, as it is likely that it is due to a spurious effect and not an extremely bright and rapid flare of \txs.

\subsubsection{Steward Observatory}
\txs was observed on 2023 June 19 UTC with the Steward Observatory 2.3-m Bok Telescope, located on Kitt Peak, AZ, using the SPOL spectropolarimeter \citep{schmidt1992}.  The faintness of the quasar dictated that the instrument be configured to provide imaging linear polarimetry \citep[see, e.g., ][]{smith2007}. A KPNO “nearly Mould” I filter was used having an effective wavelength of ~820 nm and a bandpass of $\sim185$\,nm, FWHM.  The filter bandpass selected corresponds to an effective wavelength of $\sim154$\,nm and FWHM of $\sim35$\,nm in the rest frame of the quasar.  The only major spectral feature expected within the bandpass is C IV $\lambda1549$.  The 9600-s observation yields $q = 0.0092\pm0.0161$ and $u = 0.0087\pm0.0163$ for the normalized linear Stokes parameters using a 6 arcsecond circular photometric aperture centered on \txs. As a result, only a 1-$\sigma$ upper limit of about 2.9\% can be estimated for the object’s polarization, ignoring the statistical bias associated with low-signal-to-noise linear polarization measurements.

\subsection{Effelsberg 100-m radio telescope}
We observed TXS 1508+572 over 11 months, from February 2022 to January 2023, with the Effelsberg 100-m telescope. These observations covered a wide wavelength range in four bands, centered around 60\,mm, 20\,mm, 14\,mm and 7\,mm. We performed cross-scans over the position of the point-like source in azimuth and elevation, increasing the number of repeating scans for higher frequencies to account for the lower flux densities. The data are then averaged, undergo a quality-check by a semi-automatic pipeline, corrected for pointing offsets, atmospheric opacity and elevation-dependent gain errors and lastly calibrated using 3C\,286, which is a standard calibration source. The observation and data reduction process is described in more detail in \citet{eppel2024}. We followed the same strategy, but we increased the number of scan repetitions to 32 for the highest frequencies and also included 60\,mm observations.
The radio light curves in the different bands are shown in the third panel from the bottom in Fig.~\ref{fig:MWL_LC}.
For the radio spectra, we find indices of $\alpha = -0.41\pm0.05$, $\alpha = -0.5\pm0.21$, and $\alpha = -0.24\pm0.22$ for the spectra $S\propto \nu^{\alpha}$ between 60\,mm and 20\,mm, 20\,mm and 14\,mm, and 14\,mm and 7\,mm, respectively.

In addition, for frequencies in the 60\,mm and 20\,mm band, we were able to perform polarization measurements during 13 epochs as well. We find an average polarization and standard deviation of $2.9\pm1.1$\,\%, $2.6\pm0.9$\,\% for 60\,mm and 20\,mm, respectively, and plot the changes of the polarization in the second panel from the bottom in Fig.~\ref{fig:MWL_LC}. 
In addition, we also observe some rotations in the electric vector polarization angle (EVPA), which are shown in the bottom panel of Fig.~\ref{fig:MWL_LC}.

\section{Results}

\subsection{Maximum \gm-ray flux}\label{sec:max_gammaflux}
\begin{table*}
\caption{Gamma-ray detections in different binnings and during the two brightest time ranges of the long-time flaring state, including the measured fluxes and computed luminosities. Fluxes and luminosities are given for the energy range from 100\,MeV to 300\,GeV.}
\label{tab:brightest_gammalum}
\centering
\begin{tabular}{lcccc}
\hline
Bin & TS    & Photon Index       & Flux {[}ph cm$^{-2}$ s$^{-1}${]} & Luminosity {[}erg s$^{-1}${]} \\ \hline
\multicolumn{5}{c}{MJD $59610 - 59617$}  \\ \hline
$59610 - 59617$ (7 days) & 55.54 & $2.41\pm0.21$ & $1.2\pm0.3\times10^{-7}$      & $2.6\pm1.0\times10^{49}$        \\
$59610 - 59613$ (3 days) & 37.29 &          & $1.4\pm0.4\times10^{-7}$    & $2.8\pm1.4\times10^{49}$        \\
$59611 - 59612$ (1 day)  & 24.99 &         & $2.6\pm1.0\times10^{-7}$    & $5.3\pm2.6\times10^{49}$        \\ \hline
\multicolumn{5}{c}{MJD $59806 - 59813$}              \\ \hline
$59806 - 59813$ (7 days) & 35.89 & $2.78\pm0.29$ & $1.2\pm0.3\times10^{-7}$           & $3.2\pm1.0\times10^{49}$          \\
$59809 - 59812$ (3 days) & 19.16 &             & $1.4\pm0.5\times10^{-7}$    & $3.6\pm1.5\times10^{49}$ \\
$59811 - 59812$ (1 day)  & 20.97 &             & $2.6\pm1.0\times10^{-7}$    & $6.8\pm3.0\times10^{49}$         \\ \hline
\end{tabular}
\end{table*}
\txs showed prolonged \gm-ray activity for over six months in 2022. 
During this period, the source exhibited flux variability on weekly time scales (Fig.~\ref{fig:MWL_LC}). 
In order to derive the maximum \gm-ray flux and obtain a value for the overall isotropic \gm-ray luminosity this blazar was able to produce, we compute light curves with shorter binnings that cover the two brightest bins in the 7-day binned light curve, that is, from MJD 59610 to 59616 (31 Jan 2023 -- 6 Feb 2023), and MJD 59806 to 59812 (15 Aug 2023 -- 21 Aug 2023).
In order to compute the \gm-ray luminosity from 100\,MeV to 300\,GeV, we use the spectral parameters obtained by modeling each 7-day time span. The results are given in Table~\ref{tab:brightest_gammalum}.
With values ranging from $(2.6-6.8)\times10^{49}$\,erg\,s$^{-1}$, \txs exhibits an integrated isotropic \gm-ray luminosity comparable to the most luminous blazar flares reported by the \fermi-LAT mission since 2008:
3C\,279 \citep[$\sim10^{49}$\,erg\,s$^{-1}$;][]{ackermann2016}, CTA\,102 \citep[$3\times10^{50}$\,erg\,s$^{-1}$;][]{gasparyan2018}, 3C\,454.3 \citep[$1-4\times10^{49}$\,erg\,s$^{-1}$;][]{nalewajko2013, nalewajko2017}, and 
PKS\,0402-362 \citep[$1.5\times10^{49}$\,erg\,s$^{-1}$;][]{nalewajko2013}.

\subsection{Spectral analysis}\label{sec:results_spectra}
\subsubsection{\fermi-LAT spectra}
\begin{table}[]
    \caption{Gamma-ray spectral parameters for the quiescent and flaring state for an energy range of 0.1--100\,GeV, which are also used in the broadband SED. The time ranges are from 01 Jan 2010 to (incl.) 31 Dec 2019 (MJD\,55197--59214) for the quiescent and 31 Jan to (incl.) 13 Feb 2022 (MJD\,59610--59623) for the flaring state.}
    \label{tab:fermi_spectra}
    \centering
    \begin{tabular}{lcc}
        \hline
        Parameter & Quiescent & Flare\\\hline
        TS & 79.0 & 69.4\\
        Flux \small{[$10^{-9}$\,ph\,cm$^{-2}$\,s$^{-1}$]} & $8.5\pm0.1$ & $103\pm22$ \\
        Photon Index & $2.99\pm0.13$ &  $2.36\pm0.17$\\
        Energy flux  & \multirow{2}{*}{$2.7\pm0.4$}  &  \multirow{2}{*}{$59\pm14$} \\
        \small{[$10^{-12}$\,erg\,cm$^{-2}$\,s$^{-1}$]} &   &   \\
        L$_{\gamma}$ [$10^{48}$erg\,s$^{-1}$]  &  $2.8\pm0.2$ &  $21\pm7$ \\
        \hline
    \end{tabular}
\end{table}
The \gm-ray spectra for both the quiescent and flaring states were modeled with a power law in the energy range from 100\,MeV to 300\,GeV. 
The best fit results for both spectra is listed in Table~\ref{tab:fermi_spectra}.
While for the quiescent state an integration time of 10 years was necessary to constrain the spectral parameters well, we were able to produce a spectrum with a similarly bound photon index over an integration time of only 14 days during the flaring state.
The spectrum during the flare ($\Gamma\approx2.4$) is significantly harder than the long-term quiescent state ($\Gamma\approx3$), and the flux is a factor of 12 larger.
Flaring blazars commonly show spectral hardening compared to their time-averaged states \citep[e.g.,][]{gasparyan2018}. Other high-$z$ blazars have displayed this behavior as well \citep{li2018, paliya2019}.

\subsubsection{X-ray spectra}
\begin{figure}
    \centering
    \includegraphics[width=.46\textwidth]{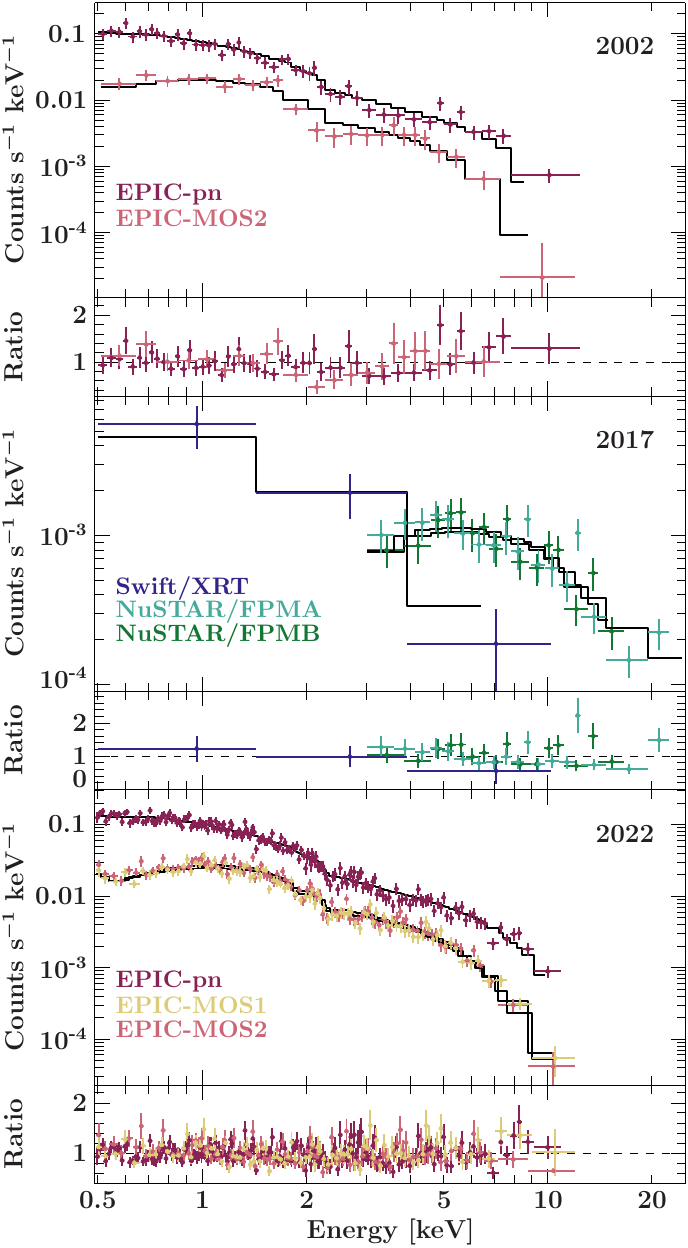}
    \caption{Simultaneous X-ray spectra taken in 2002 (top) and 2022 (bottom) with \xmm, and in 2017 with \swift/XRT and \nustar (middle), including ratios of the spectral bins versus the best-fit model.}
    \label{fig:xray_spectra}
\end{figure}

All available X-ray spectra from archival and dedicated observations are fit as described in Section~\ref{sec:xraydata}. Spectra taken with different detectors onboard \xmm are fit simultaneously, and we also fit simultaneously a \nustar and \swift/XRT observation that were taken together in April 2017. During the fit procedure, we assume that the power law index is the same across the considered full energy range and fit the flux for that entire energy range as well. In addition, we adopt a cross-normalization constant for the different detectors.
The best fit results are listed in Table~\ref{tab:xray_fitresults}, and the simultaneously fit spectra for 2002, 2017, and 2022 are plotted in Fig.~\ref{fig:xray_spectra}.
The photon statistics obtained in the \swift observations following the \gm-ray flare are insufficient to assess the presence of X-ray spectral variability. 
Comparing the two \xmm observations, one taken in 2002 and one directly after the detection of the \gm-ray flare, we find that the the X-ray flux is slightly elevated during the \gm-ray flare but the spectral shape has not significantly changed. 
The best-fit spectral index of the combined \nustar and \swift/XRT spectra taken in 2017 ($\Gamma = 1.12\pm0.07$) is harder than the \xmm spectra from 2002 and 2022, for which the spectral index can be constrained equally well.
\citet{marcotulli2020} fit the \nustar data set together with archival, non-contemporaneous data from \textit{Chandra} and \xmm (the same 2002 observation included in this work). Their simple power law fit results in a softer spectrum than found by our fit of the \nustar and \swift data together. However, the \textit{Chandra} and \xmm data sets are able to provide much better photon statistics in the soft X-ray band than the short \swift exposure, and \citet{marcotulli2020} report the finding of two breaks in their spectrum, which occur at 0.8\,keV and 6\,keV, and for which the spectrum softens in between these breaks to $\Gamma\sim1.49$, but is significantly harder at lower and higher energies ($\Gamma\sim1$). We have tested a broken power law model to search for at least the break at 6\,keV with our combined \nustar and \swift/XRT data set, but could not detect it, probably because the combined spectrum is dominated by photons detected with \nustar, for which the sensitivity only starts at 3\,keV. Hence, we cannot state whether the shape X-ray spectrum of \txs has varied over time.

\begin{table*}
\caption{Best-fit results for the X-ray spectra, fitted individually for \swift/XRT and \xmm, but for \nustar fitted combined with the simultaneous \swift/XRT observation, which is marked with $^*$ in the list of \swift observations. For \swift/XRT and \xmm, the flux is given for an energy range of 0.5--10\,keV, while the flux measured with \nustar and \swift-XRT simultaneously is given for the range 0.5--80\,keV.}\label{tab:xray_fitresults}
\resizebox{1.05\textwidth}{!}{
\begin{tabular}{llcccccccc}
\hline \hline
Instrument   & ObsID  & Date   & MJD    & Net exposure & Photon Index   & Flux  & Fit statistic vs.  & Detector       \\ 
             &        &        &        & {[}ks{]}     &                & {[}10$^{-12}$ erg cm$^{-2}$ s$^{-1}${]}  & exp. C value \& variance   &  constant                   \\ 
\hline
\multirow{5}{*}{\swift/XRT}    & 00081828001$^*$   & 2017-04-30   & 57873.4   & 2.1  & $1.17^{+0.36}_{-0.28}$   & $0.78^{+0.24}_{-0.25}$& 21.5 / $26.3\pm5.9$    & - \\
         & 00015096001  & 2022-03-28   & 59666.4  & 9.9  & $1.55\pm0.17$      & $0.68^{+0.11}_{-0.10}$& 84.1 / $96.7\pm11.4$   & - \\
         & 00015096002  & 2022-09-01   & 59823.5  & 4.8  & $1.54\pm0.20$      & $0.78^{+0.15}_{-0.13}$& 50.49 / $68.5\pm9.3$    & - \\
         & 00015096003  & 2023-03-04   & 60007.0  & 2.9  & \multirow{2}{*}{$1.29\pm0.15$}          & \multirow{2}{*}{$0.88^{+0.14}_{-0.12}$}    & \multirow{2}{*}{96.4 / $108.4\pm12.1$}       & \multirow{2}{*}{-}     \\
         & 00015096004  & 2023-03-05   & 60008.3  & 4.9  &     &  &    &   \\ \hline
\multirow{5}{*}{\xmm}          & \multirow{2}{*}{0111260201}  & \multirow{2}{*}{2002-05-11} & \multirow{2}{*}{52405.6} & 9.3 (pn)            & \multirow{2}{*}{$1.53\pm0.04$}          & \multirow{2}{*}{$0.553\pm0.022$}           & \multirow{2}{*}{829.3 / $893.9\pm39.0$}      & \multirow{2}{*}{$0.97\pm0.06$ (MOS2)}  \\
         &       &        &    & 12.1 (MOS2)         &     &  &    &   \\
         & \multirow{3}{*}{0910390101} & \multirow{3}{*}{2022-02-08} & \multirow{3}{*}{59618.1}& 58.7 (pn)           & \multirow{3}{*}{$1.481\pm0.012$}        & \multirow{3}{*}{$0.786\pm0.011$}           & \multirow{3}{*}{2729.1 / $2789.1\pm73.6$} & $1.003\pm0.020$ (MOS1) \\
         &        &        &    & 80.12 (MOS1)         &     &  &    & $1.010^{+0.020}_{-0.019}$ (MOS2)      \\
         &        &        &    & 81.1 (MOS2)         &     &  &    &   \\ \hline
\nustar & \multirow{2}{*}{60201013002} & \multirow{2}{*}{2017-04-30} & \multirow{2}{*}{57873.2} & 36.9 (FPMA)         & \multirow{2}{*}{$1.12\pm0.07$} & \multirow{2}{*}{$3.438^{+0.345}_{-0.001}$} & \multirow{2}{*}{885.9 / $943.5\pm38.3$}   & $0.97\pm0.08$ (FPMB) \\
  + \swift/XRT       &       &        &    & 36.8 (FPMB)         &     &  &    &  $0.27^{+0.07}_{-0.06}$(XRT)\\
\hline
\end{tabular}
}
\end{table*}

\subsection{Flux variability analysis}\label{sec:results_variability}
We assess the flux variability of \txs in different energy bands by computing the normalized excess variance \citep{nandra1997}, defined as 
\begin{equation}
    \sigma^2_{\mathrm{RMS}} = \frac{1}{N\mu^2}\sum^N_{i=1}[(X_i-\mu)^2-\sigma^2_i],
\end{equation}
where $N$ is the number of bins, $X_i$ is the flux or count rate in each bin, $\mu$ is the mean flux or count rate, and $\sigma_i$ is the statistical uncertainty associated with $X_i$.
Depending on the cadence of observations, we are able to examine daily to monthly time scales.
Negative values of $\sigma^2_{\mathrm{RMS}}$ indicate that no variability is present in addition to the expected noise.

\subsubsection{Short-term variability with \xmm}\label{sec:xray_variability}
\begin{figure}
    \centering
    \includegraphics[width=0.48\textwidth]{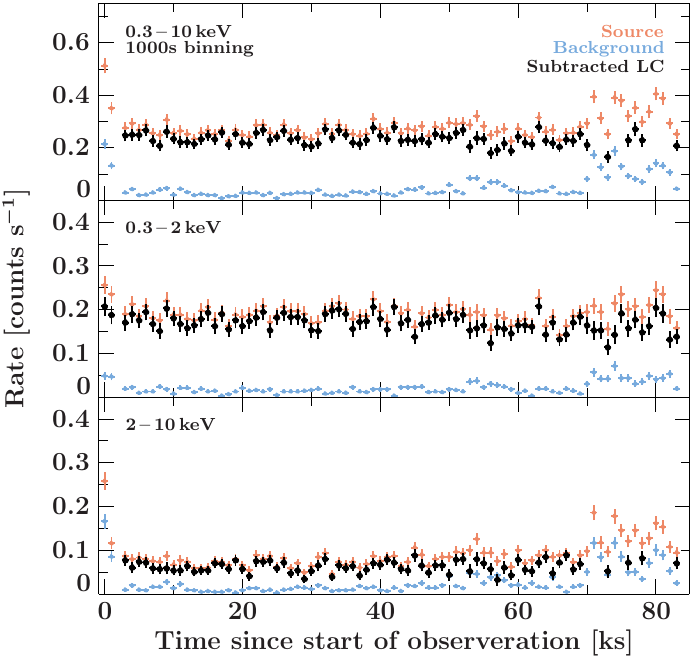}
    \caption{XMM light curve in 1\,ks binning and for different energy bands (\textit{top:} 0.3\,--\,10\,keV, \textit{middle:} 0.3\,--\,2\,keV, \textit{bottom:} 2\,--\,10\,keV). Shaded parts of the light curves mark the times in which background flaring affected the light curve in the full and hard energy range, and these times are excluded in our computation of $\sigma^2_{\mathrm{rms}}$ values.}
    \label{fig:xmm_lcs}
\end{figure}

\begin{table}[]
    \caption{Normalized excess variance in the X-rays (1\,ks binning) as observed with \xmm and in several GHz radio bands with the 100-m Effelsberg telescope.}
    \label{tab:nxs_xmm_effelsberg}
    \centering
    \begin{tabular}{lcc}
    \hline\hline
    \multicolumn{3}{c}{X-ray} \\
    Energy band & $\sigma^2_{\mathrm{RMS}}$ [$10^{-2}$] & Bins\\
    \hline
    0.3\,--\,10\,keV & $0.10\pm0.07$ & 73 \\
    0.3\,--\,2\,keV  &   $0.27\pm0.12$  & 83  \\
    2\,--\,10\,keV  &  $0.16\pm0.2$  & 71 \\
    \hline
    \multicolumn{3}{c}{Radio} \\
    Wavelength & $\sigma^2_{\mathrm{RMS}}$ [$10^{-2}$] & Bins\\
    \hline
    60\,mm  &  $-0.03\pm0.03$ & 14 \\
    20\,mm  &  $0.06\pm0.05$ & 16 \\
    14\,mm  &  $-0.13\pm0.21$ & 17 \\
    7\,mm   &  $-1.6\pm1.8$ & 10 \\
    \hline
    \end{tabular}
\end{table}

The long duration of the \xmm observation ($>80$\,ks) enables us to probe intraday flux variability of \txs at X-ray energies.  We extract binned light curves using the EPIC-pn detector with 100s-, 300s-, and 1ks-binning for the full band (0.3--10\,keV) as well as in the soft (0.3--2\,keV) and hard (2--10\,keV) band (see Section~\ref{sec:xmm_extract} for details). Parts of these light curves are affected by background flaring events, which we have filtered out by excluding periods where the background count rate lies above a certain threshold that depends on the energy band and binning.

We compute the normalized excess variance $\sigma^2_{\mathrm{RMS}}$ for all binning and energy combinations. While the values of $\sigma^2_{\mathrm{RMS}}$ for the light curves with 100s- and 300s-binning are consistent with zero (i.e., no variability detectable above the noise level) for all energy bands, the 1ks-binned light curve exhibits small source-intrinsic variability, which is strongest in the 0.3\,--\,2\,keV band (see Table~\ref{tab:nxs_xmm_effelsberg}). In the hard band, the resulting $\sigma^2_{\mathrm{RMS}}$ is consistent with zero, that is, the noise level, within uncertainties.
The 1ks-binned light curves are shown in Fig.~\ref{fig:xmm_lcs}.

\subsubsection{Flux variability in the Optical and Infrared}
\begin{figure}
    \centering
    \includegraphics[width=.48\textwidth]
    {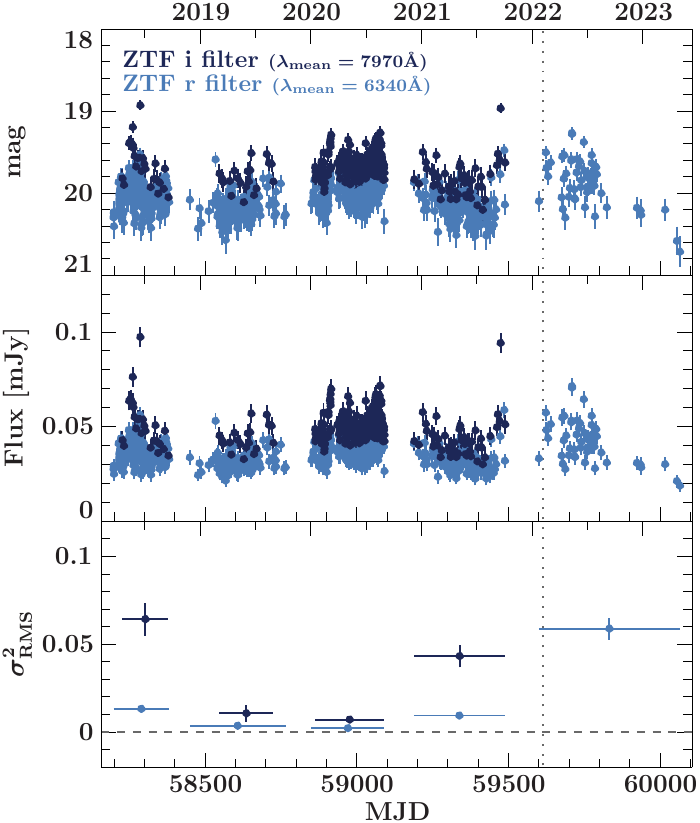}
    \caption{Optical long-term light curves in the r and i band from data taken with ZTF: light curves in magnitudes (\textit{top}), light curves in units of mJy (\textit{middle}), and normalized excess variance $\sigma^2_{\mathrm{RMS}}$ for each epoch covering a time range of roughly 10 months (\textit{bottom}). The dotted line marks the time of the flare detection at \gm-ray energies. The dashed line in the bottom panel indicates $\sigma^2_{\mathrm{RMS}}=0$.}
    \label{fig:optical_longterm_lc}
\end{figure}

\begin{figure}
    \includegraphics[width=.47\textwidth]{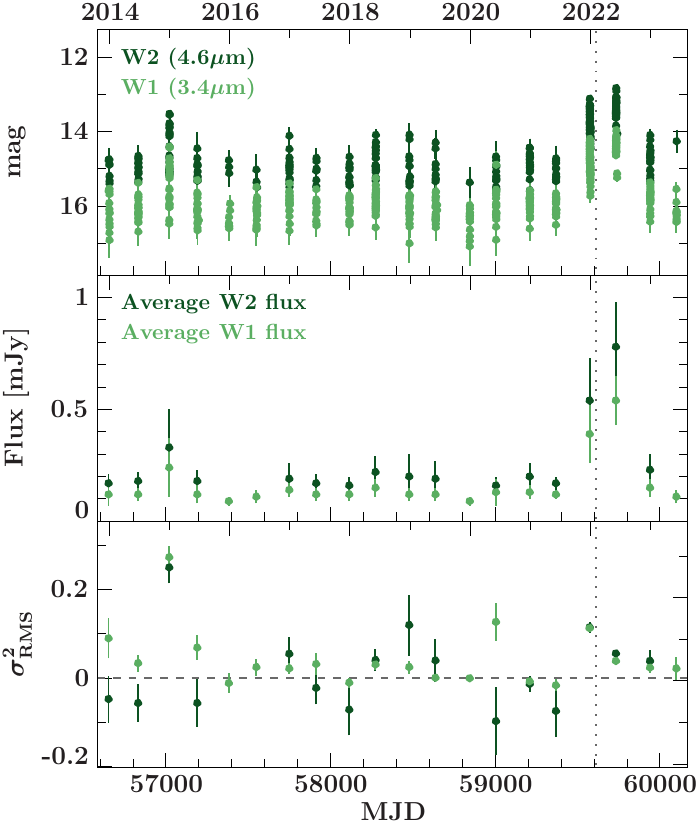}
    \caption{Infrared light curves for the W1 and W2 bands taken with {\it NEOWISE}: light curves showing all data points (\textit{top}), and the average fluxes (\textit{middle}) and normalized excess variance $\sigma^2_{\mathrm{RMS}}$ (\textit{bottom}) for each epoch from 2014 until 2023. The dotted line marks the time of the flare detection at \gm-ray energies. The dashed line in the bottom panel indicates $\sigma^2_{\mathrm{RMS}}=0$.}
    \label{fig:neowise_longterm_behaviour}
\end{figure}

At optical and infrared wavelengths, we are able to probe variability at both short- and long-term time scales. Observations were conducted from 2018 until 2023 with ZTF in the optical band (see Fig.~\ref{fig:optical_longterm_lc}), and from 2013 until 2023 with {\it NEOWISE} in the infrared band (see Fig.~\ref{fig:neowise_longterm_behaviour}). While {\it NEOWISE} covers the W1 and W2 every six months for a few days with cadences typically ranging from 90 minutes to a few hours, the monitoring cadence with ZTF varies and depends more strongly on the filter and each year.

For both the optical and infrared light curves we calculate $\sigma^2_{\mathrm{RMS}}$ for each epoch and plot them together with the light curves in Fig.~\ref{fig:optical_longterm_lc} and Fig.~\ref{fig:neowise_longterm_behaviour}, respectively. We only calculate $\sigma^2_{\mathrm{RMS}}$ for epochs that contain more than 5 bins, which is not the case for some W2 {\it NEOWISE} epochs. 
The flux variability of \txs changes over time in both the infrared and optical band. 
An infrared flare is observed to begin $\sim5$\,weeks before we reported the detection of the \gm-ray flare, peaks during the \gm-ray active phase of \txs, and gradually decreases back to baseline (see Fig~\ref{fig:neowise_longterm_behaviour}). 
Due to the sparse IR data sampling every six months, we cannot determine if the IR emission leads the \gm-ray emission since we do not know when the actual peak of the IR flux is reached. 
The strong increase in the IR flux occurs earlier than we see such a significant increase at \gm-ray energies; a rise of the \gm-ray flux with a similar amplitude would have likely caused a detection above our trigger threshold earlier on.
However, with our flare detection relying on a significant signal over 30 days prior, one could argue that there is a slight overlap between the first significant bin in the \gm-ray light curve (top panel in Fig.~\ref{fig:MWL_LC}) and the elevated IR flux measured about a week before we start to integrate over the \gm-ray flux for the first significant bin. Hence, the start of the flaring activity could have occurred simultaneously, but with the data on hand we can neither confirm nor reject it.
The short-term infrared variability peaks at the onset of the infrared flare and gradually declines back to the baseline levels. 
A similar behavior is observed in the optical data (Fig.~\ref{fig:optical_longterm_lc}).

\subsubsection{Long-term variability in the radio band}
The radio data taken with the 100-m Effelsberg telescope (light curve shown in Fig.~\ref{fig:MWL_LC}) have an average monitoring cadence of three to five weeks, depending on observing frequency.
For none of the radio bands do we find $\sigma^2_{\mathrm{RMS}}$ that is significantly above the expected noise level (see Table~\ref{tab:nxs_xmm_effelsberg}).

Interestingly, the polarization at 60\,mm is more variable, but also higher than at 20\,mm for the time range of at least 3 months before mid-2022. After that, the polarization in both bands is roughly the same apart from a dip in the 60\,mm band at MJD 59780.
Similarly, the EVPA at 60mm is rotating significantly more than the EVPA measured at 20mm.

\subsection{SED modeling}\label{sec:results_sedmodel}
\begin{figure*}
    \centering
    \includegraphics[width=.89\textwidth]{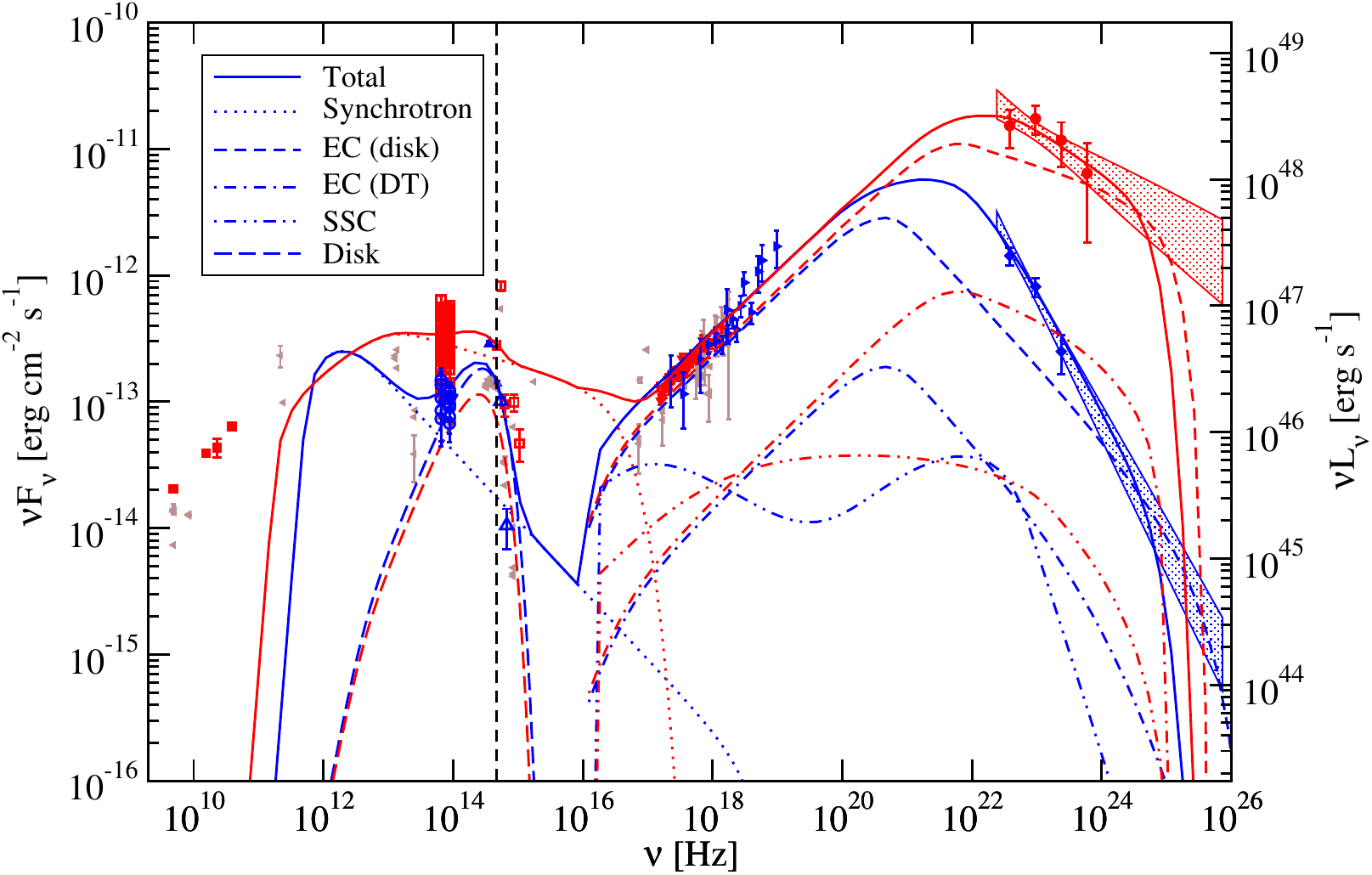}
    \caption{Broadband SED of TXS\,1508+572 in the quiescent state (blue) and during the 2022 flare (red). Archival data are taken from the online SED Builder (\url{https://tools.ssdc.asi.it/SED/})
    provided by the Space Science Data Center, and shown via gray data points. Optical data from SARA taken during the quiescent state are taken from \citet{marcotulli2020}. 
    The dashed vertical line indicates the frequency of the redshifted Ly$\alpha$ line. The model fits are produced with the code of \cite{Boettcher13}, using the parameters listed in Table \ref{tab:SEDfitparameters}. }
    \label{fig:broadbandSED}
\end{figure*}

The broadband SEDs in the quiescent and flaring states are well described by the steady-state leptonic model of \citet{Boettcher13}. It is a one-zone model, assuming a spherical emission region with radius $R_b$, located at a distance $d$ from the central supermassive black hole, moving along the jet with bulk Lorentz factor $\Gamma$, resulting in relativistic Doppler boosting of electromagnetic radiation by a Doppler factor $\delta = \left( \Gamma [ 1 - \beta_{\Gamma} \, \cos\theta_{\rm obs}] \right)^{-1}$, where $\beta_{\Gamma}$ is the normalized velocity corresponding to the Lorentz factor $\Gamma$, and $\theta_{\rm obs}$ is the angle between the jet axis and our line of sight, chosen to be $\theta_{\rm obs} = 1/\Gamma$ in order to reduce the number of free parameters. The code evaluates a self-consistent equilibrium electron distribution, based on a rapid acceleration process, injecting a power-law distribution of electrons with index $q$ between a minimum and maximum electron Lorentz factor $\gamma_{\rm min/max}$. This injection is self-consistently balanced with radiative energy losses and escape from the emission region on an escape time scale $t_{\rm esc} = \eta_{\rm esc} \, R/c$, parameterized by an escape timescale parameter $\eta_{\rm esc} \ge 1$. The resulting electron population with density spectrum $n_e (\gamma)$ in the co-moving frame of the emission region carries a power of $L_e = \pi \, R_b^2 \, \Gamma^2 \, \beta_{\Gamma} \, m_e c^3 \, \int d\gamma \gamma n_e (\gamma)$ along the jet. Radiation mechanisms included are synchrotron radiation in a tangled magnetic field of strength $B$, synchrotron-self-Compton (SSC), and external Compton scattering of both the direct accretion-disk radiation, EC (disk), and an external blackbody radiation field (temperature $T_{\rm DT}$ and radiation energy density $u_{\rm DT}$), assumed to be isotropic in the AGN rest frame, representative of infrared emission from warm dust, EC (DT). For each model simulation, the code evaluates, in addition to the electron power mentioned above, the power carried in the magnetic field (Poynting flux), $L_B = \pi \, R_b^2 \, \Gamma^2 \, \beta_{\Gamma} \, c (B^2 / 8 \pi)$, and the energy partition ratio, $\epsilon_{Be} = L_B/L_e$. 
A set of model parameters describing the quiescent and flaring SEDs 
shown in Fig. \ref{fig:broadbandSED} are listed in Table \ref{tab:SEDfitparameters}.
The chosen solutions are not necessarily unique given the significant degeneracies in the model.

Fig. \ref{fig:broadbandSED} illustrates that the SEDs in both states can be well represented with this model, with the high-energy (X-ray through $\gamma$-ray) emission having significant contributions from both EC (disk) and EC (DT), while SSC is sub-dominant. The change from the quiescent to the flaring state is achieved primarily through a harder injection spectral index of the electron spectrum and a higher $\gamma_{\rm min}$, indicating an increased electron acceleration efficiency. The optical emission, in our model fits, is dominated by the thermal accretion-disk radiation in the quiescent state, in agreement with the very low degree of polarization measured by the Steward Observatory (also in the quiescent state, although not contemporaneous with our high-energy observations). In the flaring state, the much harder electron-synchrotron spectrum dominates the optical emission. One would therefore expect significant optical polarization in this state, if the magnetic field is at least partially ordered in the emission region. Future polarimetric measurements during flaring states should be able to test this hypothesis.

We note that the jet plasma, especially in the flaring state, is out of equipartition, dominated by Poynting flux, with the chosen model parameters. The magnetic field values are also higher by a factor of
a few compared to the estimates of \cite{benke2024} for the 43~GHz core (0.8 and 1.7~G, depending on equipartition assumptions). This is consistent with the expectation that the high-energy emission region is located closer to the central engine than the 43~GHz core, as our model configuration for the high-energy emission region is still optically thick at radio frequencies. This is, however, inconsistent with the estimated distance of $0.32 \pm 0.02$~pc for the 43~GHz core. As there are significant degeneracies in our model parameters, a similarly adequate fit with a smaller value of $d$ could remedy this discrepancy (e.g., a smaller distance accompanied by an increased magnetic field). On the other hand, the estimate of the radio-core distance relied on the core-shift measurement under the assumption of a conical jet profile, and on the uncertain viewing angle. A non-conical jet profile and/or different viewing angle would obviously yield different values for the radio-core distance. 

\begin{table}
\caption{SED model fit parameters used for the modeling shown in Fig. \ref{fig:broadbandSED}.}
\centering
    \label{tab:SEDfitparameters}
\begin{tabular}{lcc}
\hline
Parameter & Quiescent & Flare \cr 
\hline 
$L_e$ [erg s$^{-1}$] & $5.6 \times 10^{45}$ & $2.7 \times 10^{45}$ \cr 
$\gamma_{\rm min}$ & 150 & 450 \cr
$\gamma_{\rm max}$ & $2.0 \times 10^5$ & $3.0 \times 10^4$ \cr 
$q$ & 3.1 & 2.3 \cr 
$B$ [G] & 5.0 & 2.8 \cr 
$\eta_{\rm esc}$ & 10 & 10 \cr 
$d$ [pc] & 0.4 & 0.35 \cr 
$\Gamma$ & 20 & 20 \cr 
$L_{\rm disk}$ [erg s$^{-1}$] & $4.0 \times 10^{47}$ & $2.5 \times 10^{47}$ \cr 
$R_b$ [cm] & $1.3 \times 10^{16}$ & $4.0 \times 10^{16}$ \cr 
$M_{\rm BH}$ [M$_{\odot}$] & $1.5 \times 10^{10}$ & $1.5 \times 10^{10}$ \cr 
$T_{\rm DT}$ [K] & $1.0 \times 10^3$ & $1.0 \times 10^3$ \cr 
$u_{\rm DT}$ [erg cm$^{-3}$] & $5.0 \times 10^{-4}$ & $5.0 \times 10^{-4}$ \cr 
\hline 
$L_B$ [erg s$^{-1}$] & $6.3 \times 10^{45}$ & $1.9 \times 10^{46}$ \cr 
$\epsilon_{Be}$ & 1.1 & 7.0 \cr 
\hline     
\end{tabular}        
\end{table}

\section{Discussion}\label{sec:discuss_conclude}
\subsection{Gamma-ray luminosity of the flare}
During the quiescent state, the \gm-ray luminosity of \txs is $2.8\times10^{48}$\,erg\,s$^{-1}$, comparable to other $z>3$ \fermi-LAT detected blazars \citep{ackermann2017}.
During the flaring state, which lasted for longer than six months in 2022 with two peak times in February and August, respectively, the \gm-ray luminosity increases to $>5\times10^{49}$\,erg\,s$^{-1}$, making this flare by \txs one of the most luminous that \fermi-LAT has ever detected.
Only two blazars, B3\,1343+451 and CTA\,102, exhibited flares with an isotropic \gm-ray luminosity $>10^{50}$\,erg\,s$^{-1}$ \citep{sahakyan2020_highz, gasparyan2018}. A few others have shown peak flare luminosities $>10^{49}$\,erg\,s$^{-1}$: 3C\,454.3 \citep{nalewajko2013, nalewajko2017}, 3C\,279 \citep{ackermann2016}, PKS\,0402$-$362 \citep{nalewajko2017}, and PKS\,0537$-$286 \citep{sahakyan2020_highz}.
Among those, B3\,1343+451 and PKS\,0537-286 are also high-$z$ objects with $z=2.53$ and $z=3.01$, respectively, showing that blazars in the early Universe are able to produce equally luminous flares; and given that three out of seven of the most extreme flares have been observed for these distant sources hints at those luminous flares being more common in high-$z$ blazars.

\subsection{Comparison of different broadband SED modeling}
The quiescent SED of \txs has been modeled by \citet{ackermann2017} using non-simultaneous data and by \citet{marcotulli2020} using a data set with simultaneous optical and X-ray data. 
All models use a one-zone leptonic model considering both SSC and EC radiation for the high-energy component, and in all cases EC dominates the X-ray and \gm-ray emission. The SEDs in \citet{ackermann2017} and \citet{marcotulli2020} include thermal emission from the dust torus even though it is not directly constrained by observational data in the infrared band. Our model does include potential radiation from the dust torus as a seed photon field for the EC component, but we find that EC from the accretion disk strongly dominates the high-energy emission.

Our SED modeling approach 
differs from that of \citet{ackermann2017} and \citet{marcotulli2020} in that we attempt to simultaneously describe the 
quiescent and flaring states of \txs while changing only a small subset of model parameters. 
Spectroscopic estimates of the black hole mass of \txs are in the range $M_\mathrm{BH}=3$--$8\times10^8$\,M$_{\odot}$ \citep{ackermann2017,diana2022}.
A standard Shakura-Sunyaev accretion disk \citep{1973A&A....24..337S} corresponding to the estimated $M_\mathrm{BH}$ values is too hot to provide the low frequency seed photons that our model favors for the origin of the high-energy SED component via the EC (disk) process. Given that the temperature-dependent peak frequency of the disk emission is $\nu_\mathrm{peak} \propto M^{-1/4}$, our model assuming a Shakura-Sunyaev disk favors a larger black hole mass of $M=1.5\times 10^{10}\,M_\odot$. 
This discrepancy is in line with quasar spectra tending to show ionization states compatible with disk temperatures lower than predicted in the standard Shakura-Sunyaev theory \citep{2013ApJ...770...30B}. 
In addition, the spin of the black hole will also have an impact on the geometry and temperature profile of the disk, which is not taken into account in our SED model. In particular, a retrograde spin would push the innermost stable circular orbit to larger distances compared to a Schwarzschild black hole \citep{1972ApJ...178..347B}, with the effect of also lowering the effective disk temperature. 

The magnetic field values favored by our SED model are also higher than those in \citet{ackermann2017} and \citet{marcotulli2020}, and we find a higher bulk Lorentz factor ($\Gamma=20$ instead of $\Gamma=11$) as well as higher $\gamma_{\mathrm{min}}$ and $\gamma_{\mathrm{max}}$ values that are expected if the electron acceleration efficiency increases during a gamma-ray flare.
The measurement of superluminal speeds $>15c$ with our VLBI monitoring campaign \citep{benke2024} supports our higher value of $\Gamma$. 
The distance of the dissipation region from the central engine is further away than described by \citet{ackermann2017}, but at a similar distance as determined by \citet{marcotulli2020}.
Earlier works by \citet{sikora2009} and \citet{ghisellini2009} argue that such a large distance favors EC emission from the broad line region or the dusty torus instead of the accretion disk due to the decrease of the energy density of disk photons and less favourable scattering geometry with increasing distance. Nonetheless, there exists currently almost no constraining information about the properties of the broad line region or the torus, and we have assumed a relatively low energy density of the isotropic external radiation field from the latter, when compared to the standard scaling relations presented in \citet{ghisellini2009}. In addition, the luminosity of the accretion disk of \txs exceeds $10^{47}$\,erg\,s$^{-1}$, and we assume a substantially larger black hole mass than what the estimates by \citet{sikora2009} and \citet{ghisellini2009} were based on. Due to this larger black-hole mass, the disk truncates and extends out to larger radii, so that IC scattering can occur under more favorable interaction angles than in the case of smaller black hole masses. The code by \citet{Boettcher13}, which is used for our modeling, takes into account the full angle dependence of the IC scattering of disk photons.

\subsection{Multiwavelength variability}
By computing the $\sigma^2_{\mathrm{RMS}}$ value for each energy band for which we have monitoring data available, albeit covering different time scales, we can compare the variability of the source at different wavelengths.
While at X-ray energies we see only a small amount of variability within one day, the IR $\sigma^2_{\mathrm{RMS}}$ values for some epochs indicate more significant variability for time ranges that cover a few days at most.
For blazars in general, the variability of IR emission seems to be correlated with the presence of \gm-ray emission \citep{mao2018}. 
The variability of \txs in the optical and IR regime seems to precede the \gm-ray activity given that already before the detection of the \gm-ray flare the IR flux and the intraday variability had increased.
Considering that \txs would be classified as a low-synchrotron-peaked (LSP) source (in the source-frame), the variability displayed at IR wavelengths would occur in the R band for a source at $z=0$. 
\citet{paliya2017} found that LSP blazars are more variable on intra-night time scales, and for which the commonly accepted scenario is that this variability originates from the non-thermal emission of the jet. Therefore, we suggest that the observed variability is connected to synchrotron emission from the jet rather than signature from the accretion disk.

\begin{figure}
    \centering
    \includegraphics[width=0.48\textwidth]{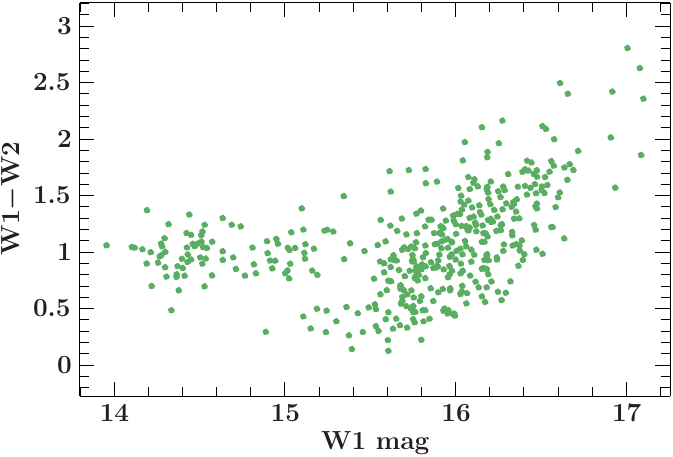}
    \caption{Change of the W1$-$W2 color with respect to the brightness of the W1 band.}
    \label{fig:color_variability_neowise}
\end{figure}

We look at possible correlations between the brightness vs. a color change, and find none at optical wavelengths. This behavior is not unusual though, as the majority of FSRQs exhibit neither a redder-when-brighter or bluer-when-brighter trend \citep{negi2022}.
In order to assess the change of the infrared W1$-$W2 color with the infrared brightness of \txs, we plot the  W1$-$W2 color over the W1 magnitude for all {\it NEOWISE} data taken simultaneously in Fig.~\ref{fig:color_variability_neowise}.
The infrared flux from \txs shows a bluer-when-brighter behavior at low fluxes and transitions to a state with no color change for $m_{\mathrm{W1}} < 15$, which could indicate the transition between a disk-dominated and synchrotron-dominated IR flux. Indeed, the bluer-when-brighter trend of IR emission has been observed with {\it NEOWISE} for the majority of both FSRQs and BL Lacs \citep{anjum2020}.

\subsection{Radio VLBI campaign and polarization}
\begin{figure*}
    \centering
    \includegraphics[width=0.95\textwidth]{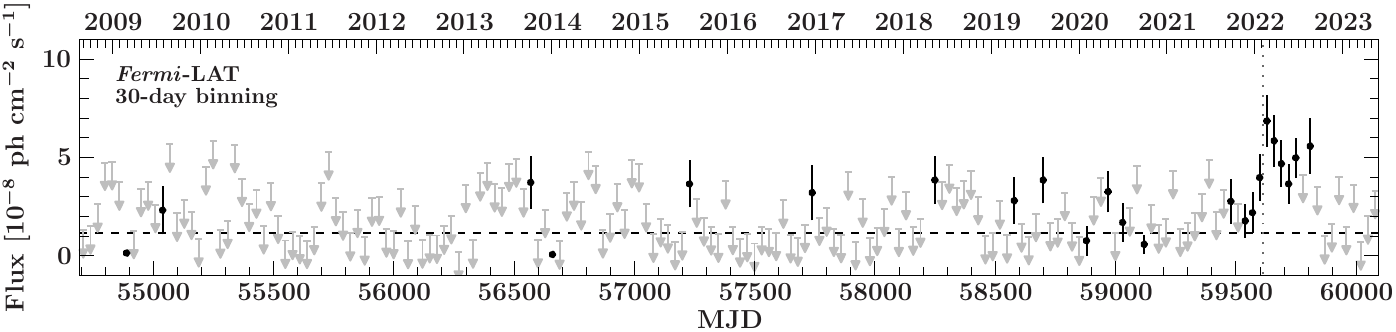}
    \caption{Long-term \fermi-LAT light curve of \txs in 30-day binning. Bins with TS$<9$ are plotted as arrows at the $2\sigma$ upper limit. The dashed horizontal line marks the longterm average flux, while the dotted vertical line marks the flare detection in 2022.}
    \label{fig:longterm_fermi_LC}
\end{figure*}
\citet{benke2024} present the results from a VLBI monitoring campaign triggered by the \gm-ray flare of \txs in 2022. The Very Long Baseline Array (VLBA) and the Effelsberg 100-m radio telescope were used to obtain milliarcsecond-scale images at 15\,GHz, 22\,GHz, and 43\,GHz. In addition, one observation at 86\,GHz was performed with the VLBA and the Green Bank Telescope, allowing to probe the jet of \txs at $456$\,GHz in the rest-frame frequency of the source.
In general, the radio observations support the underlying scenario in which the
injection of fresh electrons in the acceleration zone explains the observed broadband variability. 
By tracking the evolution of the jet over a time range of roughly 10~months, \citet{benke2024} find morphological changes in the jet and apparent superluminal speeds of the jet component motion of $\sim14c$\,--\,$32c$, depending on frequency. The latter is in agreement with the bulk Lorentz factor of $\Gamma=20$ that is favored by our SED model. 
The observed jet components can be traced back to an ejection time between 2016 and 2019, and would not be causally connected to the \gm-ray activity seen in 2022. When computing a 30-day binned light curve for \txs over the entire duration of the \fermi mission (see Fig.~\ref{fig:longterm_fermi_LC}), we find several times when the signal from the blazar is TS\,$\geq9$ ($\gtrsim3\sigma$), with four bins falling into the year 2020, which indicates enhanced activity of the source during that time range.
Assuming that the \gm-ray flare in the first half of 2022 has produced a new outflowing component, we expect to detect a new radio component to be resolvable by VLBI between 2025 and 2027.

While the VLBI radio observations spatially resolve the jet within inner several parsec of the source, the radio core itself likely consists of multiple emission zones that cannot be resolved. Radio polarization measurements at different frequencies from the Effelsberg observatory allow us to look into the structure of the unresolved radio core. 
The Effelsberg data (Fig.~\ref{fig:MWL_LC}) shows that the time variability of the polarization degree is more pronounced than the Stokes I variability, especially at 60\,mm, indicating that the core itself consists of several emission regions.

VLBI polarization measurements of \txs have been analyzed by \citet{osullivan2011}, who reported the detection of polarization at 60\,mm and 36\,mm for the core, and at 60\,mm for a single jet component. Unfortunately, our VLBI observations do not fall into a time when the 60\,mm net polarization is higher than the net polarization at 20\,mm, and without synchronous data, interpretation of the apparent short timescale variability of the polarization parameters from MJD\,59680 to MJD\,59750 (see Fig.~\ref{fig:MWL_LC}) would be too speculative.

\section{Conclusions}
The detection of a \gm-ray flare by a $z>3$ blazar is rare: PKS\,0537$-$286 is the only source which has shown multiple flares, which were so bright that they could be detected on daily time scales \citep{sahakyan2023}, and which has been communicated in real time by the LAT collaboration \citep{2017ATel10356....1C,2020ATel14285....1A,2022ATel15405....1V}.
The study by \citet{kreter2020} finds nine TS $\geq25$ detections for three $z>3$ blazars (excluding PKS\,0537$-$286) over a time span of 10 years and 8 months, resulting in a high-$z$ blazar flare every 14 months on average.

The 2022 flare from \txs reported in this paper is among the most luminous events that have been seen from this source class. 
For both the quiescent and the flaring state, the broadband SED model for \txs requires a dominant contribution from EC emission to describe the high-energy emission; similar to FSRQs in the local Universe displaying Compton dominance at all activity states \citep[e.g.,][]{krauss2016}.
Our SED modeling suggests a lower accretion disk temperature than expected in a Shakura-Sunyev disk, as well as a high bulk Lorentz factor of $\Gamma = 20$ that is in line with the superluminal motion in the jet of \txs described in \citet{benke2024}.  
At present, the data covering most of the synchrotron component are sparse, and a significant portion of optical information that could help constrain the emission from the accretion disk is lost due Lyman-$\alpha$ absorption. 

Hence, constraining the low-energy component of the SED of \txs, or for any blazar with $z>3$ for that matter, presents a challenge. The low observed optical polarization during a quiescent state allows us to conclude that the optical emission from \txs is strongly dominated by the thermal radiation from the accretion disk. However, we were not able to obtain an optical polarization measurement during the flaring state.
Similar to local FSRQs, \txs also presents variability from the jet emission, which is measured in the IR band.
While only a small number of such systems have been studied so far, current hard X-ray band and future MeV missions, such as the \textit{Compton Spectrometer and Imager} \citep[COSI;][]{tomsick2019, tomsick2023}, or the \textit{Advanced Particle-astrophysics Telescope} \citep[APT;][]{buckley2019,buckley2022}, are ideal to search for these distant objects, and their enhanced sensitivity promises more detections of powerful sources in the early Universe in coming years.


\section{Acknowledgments}
The authors express gratitude towards the anonymous referee who gave valuable comments to improve the manuscript. In addition, the authors thank Lea Marcotulli for providing SARA optical photometry measurements that are used for modeling the quiescent state SED, and David~J.~Thompson for editorial suggestions that improved the manuscript.
JH, FE, MK, and FR acknowledge support from the Deutsche Forschungs-
gemeinschaft (DFG, grants 447572188, 434448349, 465409577).
This publication is part of the M2FINDERS project which has received funding from the European Research Council (ERC) under the European Union’s Horizon 2020 Research and Innovation Programme (grant agreement No 101018682).
This research has made use of the XRT Data Analysis Software (XRTDAS) developed under the responsibility of the ASI Science Center (ASDC), Italy.
Furthermore, it made use of a collection of ISIS functions (ISISscripts) provided by ECAP/Remeis observatory and MIT (\href{https://www.sternwarte.uni-erlangen.de/isis/}{https://www.sternwarte.uni-erlangen.de/isis/}).
Part of this work is based on observations obtained with \xmm, an ESA science mission with instruments and contributions directly funded by ESA Member States and NASA.
This publication makes use of data products from the {\it Near-Earth Object Wide-field Infrared Survey Explorer (NEOWISE)}, which is a joint project of the Jet Propulsion Laboratory/California Institute of Technology and the University of Arizona. {\it NEOWISE} is funded by the National Aeronautics and Space Administration.
This research is also partly based on observations obtained with the Samuel Oschin Telescope 48-inch and the 60-inch Telescope at the Palomar Observatory as part of the Zwicky Transient Facility project. ZTF is supported by the National Science Foundation under Grant No. AST-2034437 and a collaboration including Caltech, IPAC, the Weizmann Institute for Science, the Oskar Klein Center at Stockholm University, the University of Maryland, Deutsches Elektronen-Synchrotron and Humboldt University, the TANGO Consortium of Taiwan, the University of Wisconsin at Milwaukee, Trinity College Dublin, Lawrence Livermore National Laboratories, and IN2P3, France. Operations are conducted by COO, IPAC, and UW. The ZTF forced-photometry service was funded under the Heising-Simons Foundation grant \#12540303 (PI: Graham).
In parts, this work is based on observations with the 100-m telescope of the MPIfR (Max-Planck-Institut für Radioastronomie) at Effelsberg.

\vspace{5mm}
\facilities{\fermi-LAT, \nustar, \swift/XRT, \xmm, Zwicky Transient facility, {\it NEOWISE}, 100-m Effelsberg radio telescope}

\software{fermipy \citep{fermipy},
          ISIS \citep{isis}
          }

\bibliography{txs1508}{}
\bibliographystyle{aasjournal}

\end{document}